\documentclass[10pt,conference]{IEEEtran}
\IEEEoverridecommandlockouts

\usepackage{amsmath,amssymb,amsfonts}
\usepackage{algorithmic}
\usepackage{graphicx}
\usepackage{textcomp}
\usepackage{xcolor}
\usepackage{enumitem}
\usepackage{multirow}
\usepackage{subfigure}
\usepackage{pifont}
\usepackage{mathtools}
\usepackage{color}
\usepackage{balance}

\DeclarePairedDelimiterX\braket[2]{\langle}{\rangle}{#1 \delimsize, #2}

\def\BibTeX{{\rm B\kern-.05em{\sc i\kern-.025em b}\kern-.08em
    T\kern-.1667em\lower.7ex\hbox{E}\kern-.125emX}}
\begin{document}

\title{\fontsize{24}{24} Multi-Stage Watermarking for Quantum Circuits}

\author{
\IEEEauthorblockN{Min Yang}
\IEEEauthorblockA{Indiana University Bloomington\\my36@iu.edu}
\and
\IEEEauthorblockN{Xiaolong Guo}
\IEEEauthorblockA{ Kansas State University\\guoxiaolong@k-state.edu}
\and
\IEEEauthorblockN{Lei Jiang}
\IEEEauthorblockA{Indiana University Bloomington\\jiang60@iu.edu}
}

\maketitle

\begin{abstract}

Quantum computing represents a burgeoning computational paradigm that significantly advances the resolution of contemporary intricate problems across various domains, including cryptography, chemistry, and machine learning. Quantum circuits tailored to address specific problems have emerged as critical intellectual properties (IPs) for quantum computing companies, attributing to the escalating commercial value of quantum computing. Consequently, designing watermarking schemes for quantum circuits becomes imperative to thwart malicious entities from producing unauthorized circuit replicas and unlawfully disseminating them within the market.

Unfortunately, the prevailing watermarking technique reliant on unitary matrix decomposition markedly inflates the number of 2-qubit gates and circuit depth, thereby compromising the fidelity of watermarked circuits when embedding detectable signatures into the corresponding unitary matrices. In this paper, we propose an innovative multi-stage watermarking scheme for quantum circuits, introducing additional constraints across various synthesis stages to validate the ownership of IPs. Compared to the state-of-the-art watermarking technique, our multi-stage watermarking approach demonstrates, on average, a reduction in the number of 2-qubit gates by 16\% and circuit depth by 6\%, alongside an increase in the fidelity of watermarked circuits by 8\%, while achieving a 79.4\% lower probabilistic proof of authorship.

\end{abstract}


\section{Introduction}
\label{s:intro}

Quantum computing has significantly advanced the resolution of contemporary intricate problems across various domains, including cryptography~\cite{Deutsch:PRL1996}, chemistry~\cite{Cao:CR2019}, and machine learning~\cite{Chu:QCE2023}, leveraging diverse quantum mechanisms like superposition, interference, and entanglement. The realization of a quantum application involves the construction of a quantum circuit comprising high-level complex quantum gates. Quantum circuit synthesis~\cite{Younis:QCE2021} compiles the quantum circuit into a sequence of native gates  compatible with a target Noisy Intermediate Scale Quantum (NISQ) machine. Furthermore, this synthesis process~\cite{anis2021qiskit} entails the mapping of these native gates onto the physical qubits of the NISQ machine, ultimately resulting in the conversion of this mapping into an executable schedule on the NISQ machine.

The development of a quantum circuit tailored to solve a specific problem necessitates profound domain expertise, a resource that may be lacking in small companies operating outside the realm of quantum computing. The burgeoning potential of quantum computing has spawned an emergent market scenario wherein leading quantum computing entities offer their quantum circuits as intellectual properties (IPs)~\cite{Kop:JIPLP2022} to such non-specialized small enterprises. Given the considerable commercial value attached to quantum circuit IPs, there exists a looming threat of malicious actors producing unauthorized copies~\cite{Chen:ICSM2024,Saki:ICCAD2021} and illicitly disseminating these circuits within the market. Hence, it is crucial to create a watermarking technique tailored for quantum computing firms to authenticate ownership of quantum circuit IPs, all while ensuring minimal impact on the fidelity and efficiency of the circuits executed on NISQ machines.

The state-of-the-art watermarking technique~\cite{Saravanan:ISQED2021} for quantum circuits encounters limitations due to its substantial increase in the 2-qubit gate count and the circuit depth, leading to a significant deterioration in the fidelity of watermarked circuits. Throughout the process of decomposing the unitary matrix of a quantum circuit, defined by complex high-level quantum gates, into native gates compatible with a target NISQ machine, the prior watermarking technique~\cite{Saravanan:ISQED2021} inserts a signature into the original unitary matrix of each circuit block. The signature of each circuit block must possess sufficient length to be detectable. However, the incorporation of these signatures profoundly alters the unitary matrix representing the quantum circuit, resulting in a considerable disparity between the original and watermarked unitary matrices. This pronounced unitary difference not only escalates the number of 2-qubit gates in the watermarked circuits but also amplifies the circuit depth, consequently leading to a substantial degradation in the fidelity of the watermarked circuits.

\begin{figure*}[t!]
\centering
\includegraphics[width=\textwidth]{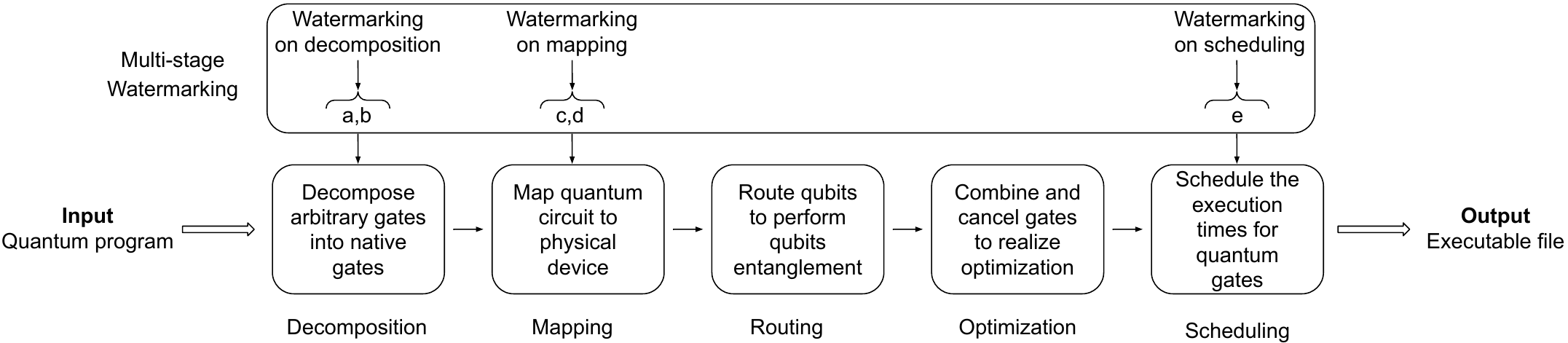}
\vspace{-0.2in}
\caption{The overview of a multi-stage watermarking for quantum circuits.}
\label{f:workflow}
\vspace{-0.1in}
\end{figure*}

In this paper, we introduce a novel multi-stage watermarking scheme tailored for quantum circuits. Diverging from the prior approach~\cite{Saravanan:ISQED2021} that solely embeds lengthy signatures during unitary matrix decomposition, our method incorporates additional constraints across various stages of quantum circuit synthesis. Leveraging an increased number of constraints, our approach enables the insertion of longer signatures with minimal modifications at each synthesis stage. Our contributions are delineated as follows:
\begin{itemize}[leftmargin=*, nosep, topsep=0pt, partopsep=0pt]
\item \textbf{Watermarking on Unitary Matrix Decomposition}. Our method adjusts the parity of the most significant digit of the distance between the original and output unitary matrices of each circuit block as a signature symbol during unitary matrix decomposition. Given a quantum circuit's multiple circuit blocks, our approach facilitates the insertion of multi-symbol signatures during decomposition.

\item \textbf{Watermarking on Qubit Mapping}. Each possible mapping between the logical qubits of a quantum circuit and the physical qubits of a NISQ machine is assigned a unique multi-symbol signature. By considering various possible mappings, our method accommodates the insertion of diverse signatures during the qubit mapping stage.

\item \textbf{Watermarking on Gate Scheduling}. During gate scheduling, our method introduces a single-symbol signature into a quantum circuit by placing two self-canceling gates before and after a SWAP gate along non-critical paths, respectively. Controlling the number of SWAP gates attached with self-canceling gates tunes the signature length.

\item \textbf{Minimal Overhead \& Enhanced IP Protection}. Compared to the state-of-the-art watermarking scheme, our technique demonstrates significant reductions in circuit depth by 6\%, decreases in the 2-qubit gate count by 16\%, and improvements in fidelity by 8\%, all while maintaining a 79.4\% lower probabilistic proof of authorship across various quantum circuit benchmarks.
\end{itemize}

\section{Background and Motivation}
\label{s:backg}

\subsection{Quantum Computing}

Despite being in the NISQ era, quantum computing has already demonstrated early quantum advantages~\cite{Arute:Nature2019,Zhong:SCIENCE2020} and shows great promise in surpassing classical computers' capabilities. Quantum computing has been applied in applications across various domains, including cryptography~\cite{Deutsch:PRL1996}, chemistry~\cite{Cao:CR2019}, and machine learning~\cite{Chu:QCE2023}. On state-of-the-art NISQ machines, such as IBM superconducting~\cite{Gambetta:IBM2020} or Quantinuum Trapped-Ion~\cite{Moses:PRX2023} machines, a program is realized through a quantum circuit composed of high-level complex quantum gates, which are designer-friendly. As Figure~\ref{f:workflow} shows, quantum circuit synthesis~\cite{Younis:QCE2021} transpiles the quantum circuit into a sequence of native gates supported by the target NISQ machine. Subsequently, the synthesis procedure~\cite{anis2021qiskit} maps the native gates to physical qubits on the NISQ machine and generates an executable schedule for these gates.

\subsection{Intellectual Property: Quantum Circuit}

Developing effective quantum circuits tailored to solve specific problems demands profound expertise in quantum mechanics, algorithms, and computational complexity. However, such expertise may be scarce in smaller companies or organizations not specialized in quantum computing. With the advancement of quantum computing, there is a burgeoning market for quantum circuit designs. Established quantum computing entities are now offering their circuits as IPs~\cite{Kop:JIPLP2022} to smaller enterprises lacking in-house quantum expertise. Similar to classical computing, there arises a concern about unauthorized copying and dissemination of these circuits by malicious actors~\cite{Saki:ICCAD2021,Chen:ICSM2024}, posing a threat to the commercial interests of quantum computing firms. To mitigate the risk of unauthorized copying, the adoption of watermarking techniques designed for quantum circuits has become imperative.

\subsection{Quantum Circuit Synthesis}

As illustrated in Figure~\ref{f:workflow}, the quantum circuit synthesis process primarily encompasses five steps~\cite{Younis:QCE2021,anis2021qiskit,bqskit}. Initially, the high-level complex quantum gates within a quantum circuit are decomposed into a sequence of native gates compatible with a target NISQ machine. Subsequently, the logic qubits within the circuit are assigned to the physical qubits of the target NISQ machine~\cite{tannu2019not}. Next, SWAP gates are introduced to facilitate the routing of logic qubits until two logic qubits become adjacent before the imminent two-qubit gate operation on them. Each SWAP gate represents a costly and noisy operation~\cite{li2019tackling}. Following that, various circuit optimizations are applied, such as merging consecutive gates, eliminating redundant gates~\cite{maslov2008quantum}, and identifying shorter gate sequences, to reduce the circuit complexity while preserving its intended functionality. Finally, all quantum gates in the circuit are scheduled to minimize program runtime or maximize fidelity.

\subsection{Classic CMOS Hardware Watermarking}

Similar to quantum circuits, preserving the IPs of classical CMOS circuits is paramount through watermarking in CMOS logic synthesis~\cite{Anandakumar:CEA2024}. CMOS circuit synthesis involves tackling an NP-hard optimization problem~\cite{Sengupta:TCAD2018}, where exhaustively exploring all potential solutions is impractical. Typically, heuristic algorithms are employed to search for near-optimal solutions by traversing the vast design space while adhering to design constraints. To watermark a CMOS circuit, the authorship message is initially translated into a set of embedded constraints~\cite{Anandakumar:CEA2024}. These embedded constraints serve as additional inputs to the heuristic algorithm. The resulting near-optimal solution obtained by the algorithm yields a watermarked CMOS IP circuit that satisfies both the original design constraints and the embedded watermarking constraints. Previous studies~\cite{Sengupta:TCAD2018,Anandakumar:CEA2024} define the probabilistic proof of authorship (PPA) to measure the likelihood of an actor successfully satisfying the final solution to the IP design watermarking problem. A lower PPA indicates better IP protection for the watermarked circuits. Further details on PPA can be found in Section~\ref{s:exp}.

\section{Related Work and Motivation}
\label{s:related}

The previous quantum circuit watermarking technique~\cite{Saravanan:ISQED2021} directly integrates signatures into the unitary matrix representing a quantum circuit during the unitary matrix decomposition stage, i.e., the initial step in quantum circuit synthesis. However, to ensure detectability, these signatures must contain numerous bits and be sufficiently lengthy, necessitating significant modifications to the unitary matrix of the quantum circuit. This substantial disparity between the original and watermarked unitary matrices of the quantum circuit notably increases the number of 2-qubit gates and the circuit depth, resulting in a noticeable degradation in the fidelity of the watermarked circuit. We evaluate the fidelity of a quantum circuit using the Probability of Successful Trials (PST)~\cite{tannu2019not}, elaborated in Section~\ref{s:exp}, which outlines our experimental methodology. A higher PST indicates a higher fidelity for the quantum circuit. Table~\ref{t:water_moti_example} demonstrates that the previous technique significantly increases circuit depth and the number of 2-qubit CNOT gates by $9\%\sim29\%$ and $9.3\%\sim31\%$, respectively, across various quantum circuits, thereby reducing their PST values by $2\%\sim11\%$ through the insertion of a 2-bit signature into each circuit block. To address this challenge, we propose a novel multi-stage watermarking technique, imposing additional design constraints without significantly increasing circuit depth or the number of 2-qubit gates. Our approach aims to minimize the negative impact on fidelity across various quantum applications.

\begin{table}[t!]
\centering
\caption{The design overhead of the state-of-the-art watermarking scheme~\cite{Saravanan:ISQED2021} for quantum circuits.}
\label{comparison_motivation}
\setlength{\tabcolsep}{3pt}
\begin{tabular}{|c|c|c|c|c|c|c|} 
\hline
\multirow{3}{*}{Benchmark} & \multicolumn{3}{c|}{no watermark} & \multicolumn{3}{c|}{\cite{Saravanan:ISQED2021}}        \\\cline{2-7}
                           & circuit  & CNOT     & \multirow{2}{*}{PST}      & circuit  & CNOT     & \multirow{2}{*}{PST}\\
												   & depth    & gate \#  &                           & depth    & gate \#  &                       \\\hline\hline
fredkin\_n3                & 31       & 16       & 0.8281                    & 40       & 21       & 0.8032              \\\hline
decod24-v2\_43             & 63       & 32       & 0.8027                    & 69       & 35       & 0.7231              \\\hline
4gt11\_84                  & 34       & 17       & 0.8505                    & 32       & 17       & 0.8098                \\\hline
\end{tabular}
\vspace{-0.1in}
\label{t:water_moti_example}
\end{table}

\section{Multi-Stage Watermarking}
\label{s:mutli-stage}

In this paper, we present a novel constraint-based multi-stage watermarking technique for quantum circuits. Figure~\ref{f:workflow} illustrates an overview of our proposed technique integrated into the whole quantum circuit synthesis process. The synthesis process initiates with a quantum circuit and a signature message. The quantum circuit is typically written in OpenQASM, while the signature message, only available to the IP owner, consists of multiple symbols. Our multi-stage watermarking approach translates each symbol in the signature message into a embedding constant, satisfying all embedding constants at various stages of the quantum circuit synthesis process, including decomposition, mapping, and scheduling. Consequently, the outcome of this synthesis process is the watermarked and synthesized quantum circuit, directly executable on the target NISQ machine.

\begin{figure}[t!]
  \centering
  \includegraphics[width=0.8\linewidth]{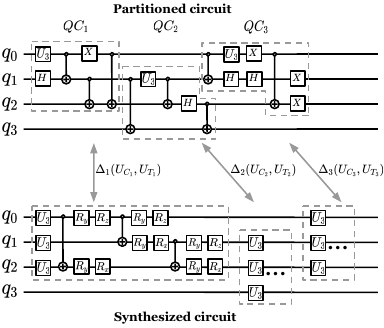}
	\vspace{-0.1in}
  \caption{Stage-1: watermarking during unitary matrix decomposition and native gate translation.}
  \label{stage1_example}
	\vspace{-0.2in}
\end{figure}

\subsection{Stage-1: Watermarking on Unitary Matrix Decomposition}
\label{s:UMD}

\begin{figure*}[t!]
\centering
\vspace{-0.3in}
\subfigure[The backend and its all mappings.]{
   \includegraphics[width = 2.2in]{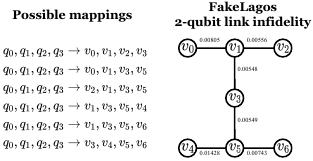}
   \label{f:stage2_example1}
}
\subfigure[Mapping-based watermarking.]{
   \includegraphics[width=4.6in]{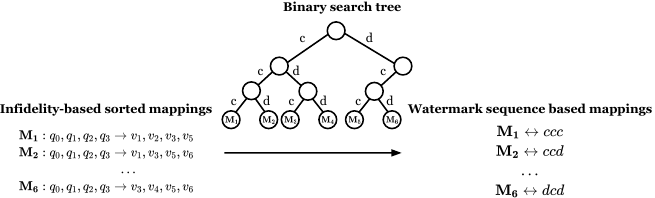}
   \label{f:stage2_example2}
}
\caption{Stage 2: Watermarking on qubit mapping.}
\label{stage2_example}
\vspace{-0.2in}
\end{figure*}

\textbf{Decomposition Process}. A quantum circuit is denoted by $(Q, G)$, where $Q$ represents the set of logical qubits in the circuit, and $G$ indicates the set of quantum gates to be applied to these qubits. For an $n$-qubit quantum circuit, the unitary matrix decomposition process involves breaking down its unitary matrix $U_{T}$ into a sequence of smaller unitary matrices. Each of these matrices can be implemented by native quantum gates supported by the target NISQ machine. The quality of the output of unitary matrix decomposition is assessed using the distance ($\Delta(U_{C}, U_{T})$) between the original unitary matrix $U_{T}$ and the output unitary matrix $U_{C}$. Typically, this distance should be smaller than a predefined threshold $\epsilon$:
\begin{equation}
\Delta(U_{C}, U_{T}) = 1 - \frac{|\text{Tr}(U_{C}^{\dagger}U_{T})|}{2^n}<\epsilon,
\end{equation}
where $\text{Tr}$ represents the Hilbert-Schmidt inner-product operation, and $n$ is the number of logical qubits. However, the unitary matrix of a quantum circuit can be excessively large, with a size of $2^n\times2^n$, exponentially increasing with $n$. To decompose such a large unitary matrix, the following steps are required:
\begin{itemize}[leftmargin=*, nosep, topsep=0pt, partopsep=0pt]

\item \textit{Partition}: The large quantum circuit is partitioned into $k$ smaller sub-circuits $QC_i = (Q_i, G_i)$, where each $Q_i \subseteq Q$, each $G_i \subseteq G$, and $i\in[1,k]$. Importantly, there should be no overlap between sub-circuits. The partitioning is expressed as:
\begin{equation}
QC = \bigoplus_{i=1}^{k} QC_i,
\end{equation}
where $\bigoplus$ signifies the composition operation for combining sub-circuits to construct the complete quantum circuit.

\item \textit{Decompose}: Each sub-circuit $QC_i$ undergoes independent decomposition to yield a unitary matrix $U_{C_i}$. By deleting or adding few 2-qubit gates, the approximate decomposition scheme $S$~\cite{patel2022quest} is applied to each sub-circuit $QC_i$ to generate its corresponding output unitary matrix $S(QC_i)=U_{C_i}$, and controls the distance between original and target unitary matrices. The decomposition must satisfy:
\begin{equation}
\Delta_i(U_{C_i}, U_{T_i}) < \epsilon_i,    
\end{equation}
where $U_{T_i}$ is the target unitary for sub-circuit $QC_i$, $\Delta_i(U_{C_i}, U_{T_i})$ represents the distance between $U_{C_i}$ and $U_{T_i}$, and $\epsilon_i$ is the error tolerance for that sub-circuit.

\item \textit{Assemble}: After decomposing each sub-circuit, the decomposed sub-circuits $U_{C_i}$ are combined to construct the output unitary $U_C$, depicted as:
\begin{equation}
U_C = \bigotimes_{i=1}^{k} U_{C_i},
\end{equation}
where $\bigotimes$ denotes the tensor product, combining the unitary matrices of sub-circuits into a single output unitary matrix. The assembled unitary $U_C$ should closely approximate the target unitary $U_T$ within the global error tolerance $\epsilon$.

\end{itemize}

\textbf{Watermarking Process}. During the decomposition process, our multi-stage watermarking scheme utilizes the parity of the most significant digit of the distance $\Delta_i(U_{C_i}, U_{T_i})$ between the target and output unitary matrices for each sub-circuit as a signature symbol. For each sub-circuit $QC_i$, the parity of the most significant digit of the distance $\Delta_i(U_{C_i}, U_{T_i})$ is determined according to the following rules:
\begin{enumerate}[leftmargin=*, nosep, topsep=0pt, partopsep=0pt]
\item If a signature symbol `a' is applied to $QC_i$, our technique tunes the most significant digit of $\Delta_i(U_{C_i}, U_{T_i})$ to an odd number, specifically 1, 3, 5, 7, or 9.
\item If a signature symbol `b' is inserted into $QC_i$, our watermarking technique adjusts the most significant digit of $\Delta_i(U_{C_i}, U_{T_i})$ to an even number including 2, 4, 6, or 8.
\end{enumerate}

\textbf{Watermarking Example}. Figure~\ref{stage1_example} highlights an example of our stage-1 watermarking technique applied to a quantum circuit, where a 4-qubit circuit is partitioned into three 3-qubit sub-circuits $QC_1$, $QC_2$, and $QC_3$. Our watermarking technique embeds three signature symbols into $QC_1$, $QC_2$, and $QC_3$, respectively. Specifically, the technique adjusts the parity of the most significant digit of the distance $\Delta_i(U_{C_i}, U_{T_i})$ between the target and output unitary matrices for each sub-circuit as a signature symbol. If the signature symbol for $QC_1$ is `a', the parity of the most significant digit of $\Delta_1$ is odd; otherwise, it is even. Given the signature symbols `aba' for $QC_1$, $QC_2$, and $QC_3$ respectively, the parities of the most significant digits of $\Delta_1$, $\Delta_2$, and $\Delta_3$ are odd, even, and odd.

\begin{table}[t!]
\centering
\caption{The stage-1 result.}
\label{table_stage1}
\begin{tabular}{|c||c|c|c|} 
\hline
Signature     & Circuit Depth   & CNOT Gate \#  & PST     \\\hline\hline
aa (baseline) & 55              & 32            & 0.8007  \\\hline
ab            & 61              & 32            & 0.8222   \\\hline
ba            & 63              & 33            & 0.8027   \\\hline
bb            & 57              & 33            & 0.7952  \\\hline
\end{tabular}
\vspace{-0.2in}
\end{table}

\textbf{Watermarking Result}. We applied our stage-1 watermarking technique to the benchmark circuit `decod24-v2\_43', which can be partitioned into two 3-qubit sub-circuits for unitary matrix decomposition. Our approach may potentially insert four types of signatures into these two sub-circuits during the decomposition stage. Table~\ref{table_stage1} presents the circuit depth, the number of 2-qubit CNOT gates, and the PST values for each signature sequence. Typically, the parities of the most significant digits of $\Delta_1$ and $\Delta_2$ are 1s, making `aa' become our non-watermarking baseline. Compared to this baseline, the signature insertion slightly increases the circuit depth, with the `ba' sequence showing the most significant increase. The count of CNOT gates remains relatively stable, except for the `ba' and `bb' signature, which slightly increase the count to 33. Notably, the `ab' sequence enhances the PST to 0.8222, the highest among the variations, suggesting potential fidelity improvements. These findings highlight the nuanced impact of our stage-1 watermarking on quantum circuits.

\begin{figure*}[t!]
\centering
\vspace{-0.1in}
\subfigure[A quantum circuit.]{
   \includegraphics[width=1.8in]{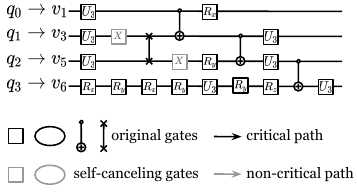}
   \label{f:stage3_critical_example}
}
\subfigure[The DAG w/o watermark.]{
   \includegraphics[width=2.3in]{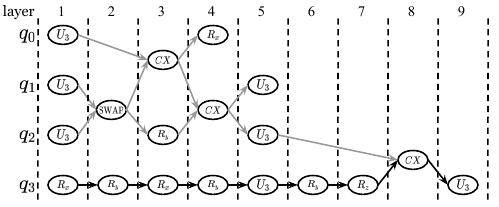}
   \label{f:freq_tech21_dis}
}
\subfigure[The DAG with `e'.]{
   \includegraphics[width=2.3in]{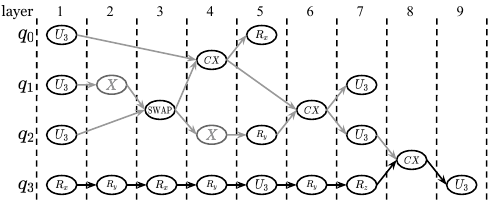}
   \label{f:freq_tech22_dis}
}
\vspace{-0.1in}
\caption{Stage-3: Watermarking on gate scheduling.}
\label{stage3_example3}
\vspace{-0.1in}
\end{figure*}

\subsection{Stage-2: Watermarking on Qubit Mapping}

\textbf{Qubit Mapping}. As illustrated in Figure~\ref{f:stage2_example1}, on a NISQ machine, the connectivity between physical qubits is modeled by an undirected graph $AG = (V, E)$~\cite{tannu2019not}, where $V$ represents the set of vertices corresponding to physical qubits, and each pair $\{v_i, v_j\} \in E$ denotes a bidirectional connection between physical qubits $v_i$ and $v_j$. When two physical qubits are connected bidirectionally, two-qubit operations such as CNOT gates can be executed between them. The qubit mapping process~\cite{tannu2019not} assigns each logical qubit $q \in Q$ in a quantum circuit to a physical qubit $v \in V$ on the NISQ machine, ensuring that $|Q|\leq |V|$. A mapping $M_i: \{q_0,q_1,...,q_N\} \rightarrow \{v_{x_0}, v_{x_1},..., v_{x_N}\}$ represents an assignment of logical qubits from the quantum circuit onto a subset of physical qubits $\{v_{x_0}, v_{x_1},..., v_{x_N}\} \subset V$. When $|Q|<|V|$, there exist multiple possible mappings between logical and physical qubits. The infidelity overhead~\cite{tannu2019not} introduced by each possible mapping can be computed by multiplying the infidelity values of 2-qubit CNOT gates between all included physical qubits. As Figure~\ref{f:stage2_example2} shows, all possible mappings are then ranked by their infidelity overhead, resulting in an ordered set $\{M_1, M_2, ..., M_N\}$, where the infidelity overhead of $M_i$ is less than that for $M_{i+1}$. Typically, the mapping~\cite{tannu2019not} introducing the lowest infidelity overhead is selected at the end of the qubit mapping stage.

\textbf{Watermarking Process}. During the qubit mapping stage, our multi-stage watermarking technique constructs an ordered binary search tree for all possible qubit mappings. The leaf nodes of the tree represent qubit mappings sorted by their infidelity overhead, as depicted in Figure~\ref{f:stage2_example2}. Our approach assigns signature symbols to the edges in the tree, using `c' for the left child and `d' for the right child. We derive a multi-symbol signature on the qubit mapping represented by a leaf node by following the symbols along the path from the root to that leaf node. For instance, the signature of $M_1$ is `ccc'. This method ensures that each mapping $M_i$ is associated with a unique signature.

\begin{table}[t!]
\centering
\caption{The stage-2 result.} 
\label{table_stage2}
\begin{tabular}{|c||c|c|c|} 
\hline
Signature       & Circuit Depth & CNOT Gate \# & PST\\  \hline\hline
ccc (baseline)  & 63            & 32           & 0.8027  \\ \hline
ccd             & 63            & 32           & 0.788   \\ \hline
cdc             & 60            & 35           & 0.7978  \\ \hline
cdd             & 61            & 29           & 0.7910  \\ \hline
dcc             & 60            & 32           & 0.7668  \\ \hline
dcd             & 60            & 29           & 0.7607  \\\hline
\end{tabular}
\vspace{-0.2in}
\end{table}

\textbf{Watermarking Example}. An illustration of watermarking on qubit mapping is depicted in Figure~\ref{stage2_example}, where we employ the quantum circuit discussed in the example in Section~\ref{s:UMD} and map it onto an IBM NISQ machine simulator known as FakeLagos, simulated using Qiskit~\cite{anis2021qiskit}. FakeLagos is represented by an undirected graph $G = (V, E)$, where $V = \{v_0, v_1, v_2, v_3, v_4, v_5, v_6\}$ denotes the set of vertices, each representing a physical qubit on the NISQ machine, and $E$ is the set of edges between vertices. Each edge indicates a link between two physical qubits, with its weight corresponding to the infidelity of 2-qubit gates on the link. For example, the edge connecting $v_0$ and $v_1$ has a 2-qubit gate infidelity of 0.00805. The quantum circuit consists of four logical qubits, $q_0,\ldots,q_3$, and a viable mapping identifies a 4-node connected sub-graph in $G$ containing all four logical qubits. Six potential mappings are identified, sorted by their infidelity values as $\{M_1, M_2, ..., M_6\}$ in ascending order. $M_1$ exhibits the lowest infidelity, whereas $M_6$ has the highest. These mappings are further organized into an ordered binary search tree, with leaf nodes corresponding to $\{M_1, M_2, ..., M_6\}$. Following the labeling of the edges of the tree with `c' and `d', each qubit mapping is associated with a unique signature. For instance, during the qubit mapping stage, the signature `ccd' uniquely identifies the mapping of the logical qubits of the quantum circuit onto the physical qubits ${v_1, v_3, v_5, v_6}$.

\textbf{Watermarking Result}. Mapping the quantum circuit benchmark `decod24-v2\_43' onto FakeLagos yields 6 possible mappings, each represented by a unique signature consisting of `c's or `d's. Table~\ref{table_stage2} illustrates the design overhead of different signatures on the mapping of `decod24-v2\_43' on FakeLagos, revealing the nuanced impacts of our watermarking technique on the quantum circuit. The insertion of a signature in the qubit mapping stage does not uniformly affect the circuit depth, with signatures like `cdc', `dcc', and `dcd' reducing the circuit depth to 60. The variation in the CNOT gate count is evident, with `cdc' increasing the count to 35, while `cdd' and `dcd' decrease it to 29. PST shows slight reductions across all signatures, with `dcc' and `dcd' exhibiting the most significant drops to 0.7668 and 0.7607, respectively. This minor decrement in PST underscores the potential negative impact of our watermarking technique. The gradual decline in PST is attributed to the increasing total infidelity as the leaf nodes of the binary tree progressively move to the right.

\subsection{Stage 3: Watermarking on Gate Scheduling}

\textbf{Gate Scheduling}. Following the mapping of logical qubits from a quantum circuit to the physical qubits of a NISQ machine, additional SWAP gates~\cite{Wille:ASPDAC2014} are introduced into the gate schedule of the quantum circuit to accommodate the physical constraints of the NISQ machine. These SWAP gates ensure that the two qubits involved in each 2-qubit gate are properly connected before the gate operation. During the gate scheduling stage, the quantum circuit is represented as a directed acyclic graph (DAG), with the SWAP gates inserted into the DAG facilitating entanglements between non-adjacent physical qubits. Within the DAG, the critical path of the quantum circuit delineates the longest sequence of quantum gates, thereby determining the circuit depth. Additional quantum gates introduced along non-critical paths of the circuit typically do not impact the circuit depth. During gate scheduling, our multi-stage watermarking technique incorporates a pair of self-canceling gates (e.g., $X$ gates) before and after some SWAP gates along the non-critical paths of a quantum circuit.

\textbf{Watermarking Process}. In the gate scheduling stage, our multi-stage watermarking technique examines all SWAP gates along the non-critical paths of a quantum circuit. A symbol `e' is employed to denote a pair of self-canceling gates introduced before and after each SWAP gate. The number of `e' in a signature indicates the number of pairs of self-canceling gates inserted along the non-critical paths of the quantum circuit. For instance, the signature `ee' signifies the addition of two pairs of $X$ gates both before and after two SWAP gates along the non-critical paths. The maximum count of `e' in a signature corresponds to the total number of SWAP gates encountered along the non-critical paths of the quantum circuit.

\textbf{Watermarking Example}. Figure~\ref{stage3_example3} illustrates an instance of our watermarking technique applied to the gate scheduling stage of a quantum circuit. The architecture of the quantum circuit is depicted in Figure~\ref{f:stage3_critical_example}, where a pair of $X$ gates is incorporated before and after the SWAP gate between $q_1$ and $q_2$. Original gates in the quantum circuit are denoted in black, while the newly added $X$ gates are represented in gray. The DAG of the quantum circuit is presented in Figure~\ref{f:freq_tech21_dis}, where the critical path is indicated by back arrows, and the non-critical paths are depicted with gray arrows. The circuit depth is 9. The DAG of the quantum circuit watermarked by our technique with `e' in the gate scheduling stage is displayed in Figure~\ref{f:freq_tech22_dis}. In this example, the inclusion of the `e' signature does not augment the circuit depth. Nonetheless, in some scenarios, our approach in the gate scheduling stage might marginally increase the circuit depth of the watermarked quantum circuits.

\begin{table}[t!]
\centering
\caption{The stage-3 result.}
\label{table_stage3}
\begin{tabular}{|c||c|c|c|} 
\hline
Signature     & Circuit Depth & CNOT Gate \# & PST    \\ \hline\hline
baseline      & 63            & 32           & 0.8027       \\ \hline
e             & 64            & 32           & 0.8076       \\ \hline
ee            & 65            & 35           & 0.7890       \\ \hline
eee           & 66            & 32           & 0.8164       \\ \hline
eeee          & 67            & 32           & 0.7949       \\\hline
\end{tabular}
\vspace{-0.2in}
\end{table}

\textbf{Watermarking Result}. Following the qubit mapping stage, the quantum circuit benchmark `decod24-v2\_43' contains two SWAP gates along the non-critical paths and five SWAP gates along the critical path. Our watermarking technique introduces one or multiple pairs of $X$ gates before and after the SWAP gates along its non-critical paths. The symbol number of the signature in this stage corresponds to the number of pairs of $X$ gates added on the non-critical paths. Table~\ref{table_stage3} presents the circuit depth, the count of CNOT gates, and the PST values of the watermarked `decod24-v2\_43' with various signatures. In this benchmark, an increasing length of the signature during our stage-3 watermarking progressively increases the circuit depth. However, the count of CNOT gates fluctuates, peaking at 35 with the `ee' signature, before reverting back to our baseline's value of 32 with further increases in the signature length. The PST value undergoes slight variations, reaching a maximum of 0.8164 with the `eee' signature, indicating the nuanced impact of our watermarking on the circuit fidelity.

\section{Experimental Methodology}
\label{s:exp}

\textbf{Baseline}. The only existing watermarking technique for quantum circuits, to the best of our knowledge, is proposed in~\cite{Saravanan:ISQED2021}, which we adopted as our baseline. In this prior technique, signatures are inserted into the original unitary matrix representing a quantum circuit during the unitary matrix decomposition stage. To ensure detectability, a significant disparity between the original and target unitary matrices is required, leading to considerable design overhead, including increased circuit depth, extra 2-qubit gates, and decreased PST values. Additionally, we consider the synthesized quantum circuits with signatures containing all `a's, all `c's, and no `e's as our non-watermarked baseline.

\begin{table}[t!]
\centering
\caption{The simulated quantum circuit benchmarks.}
\label{details_benchmarks}
\begin{tabular}{|c||c|c|c|} 
\hline
Benchmark      & Qubit \# & Circuit Depth & CNOT Gate \# \\ 
\hline\hline
fredkin\_n3    & 3      & 11    & 8          \\ 
\hline
ex-1\_166      & 3      & 12    & 9          \\ 
\hline
decod24-v2\_43 & 4      & 30    & 22         \\ 
\hline
rd32-v1\_68    & 4      & 21    & 16         \\ 
\hline
alu-v0\_27     & 5      & 21    & 17         \\ 
\hline
4gt11\_84      & 5      & 11    & 9          \\
\hline
\end{tabular}
\vspace{-0.2in}
\end{table}

\textbf{Evaluation Metrics}. To efficiently evaluate the fidelity overhead of our multi-stage watermarking scheme, we adopt the ``Probability of Successful Trials'' (PST)~\cite{tannu2019not}, which exhibits a strong positive correlation~\cite{wang2022quest} with the fidelity of a quantum circuit. We measure the difference in the PST of a quantum circuit before and after applying our watermarking technique. PST of a quantum circuit is defined as:
\begin{equation}
PST = \frac{num_{trials=initial}}{num_{trial}},
\label{e:water_pst_def}
\end{equation}
where $num_{trials=initial}$ represents the number of trials with an output identical to that of the initial state, and $num_{trial}$ denotes the total number of trials. Additionally, we consider differences in circuit depth and the number of 2-qubit CNOT gates before and after applying our watermarking. We report the distance between the original and target unitary matrices of each quantum circuit as auxiliary data to compare circuit fidelity. To measure the detectability of different watermarking techniques, we utilize the ``Probabilistic Proof of Authorship'' (PPA)~\cite{Sengupta:TCAD2018,Anandakumar:CEA2024}, defined as:
\begin{equation}
PPA=P(x\leq b)=\sum_{i=0}^b[(C!/(C-i)!\cdot i!) \cdot p^{C-i}\cdot (1-p)^i],
\label{e:eq_owner_all}
\end{equation}
where $p$ represents the probability of coincidentally satisfying a single random constraint, $PPA$ denotes the proof of authorship, $b$ signifies the number of unsatisfied constraints, and $C$ indicates the total number of imposed constraints. When employing constraint-based watermarking strategies, minimizing $PPA$ is crucial to ensure that only the author can successfully satisfy the final solution to the IP design watermarking problem. Lastly, we measure the time difference of the entire quantum circuit synthesis process with and without our watermarking technique to evaluate its time complexity.

\begin{figure*}[t!]
\centering
\subfigure[Circuit depth.]{
   \includegraphics[width=3.3in]{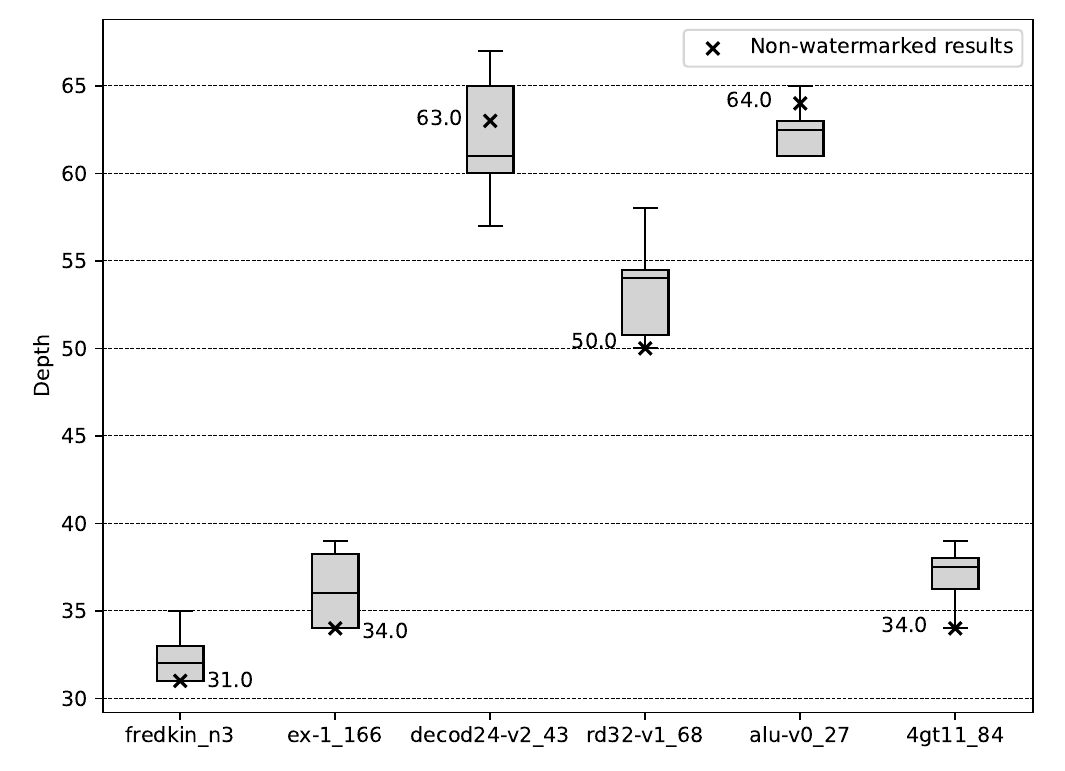}
   \label{f:Comparative1}
}
\hspace{-0.2in}
\subfigure[CNOT gate \#.]{
   \includegraphics[width=3.3in]{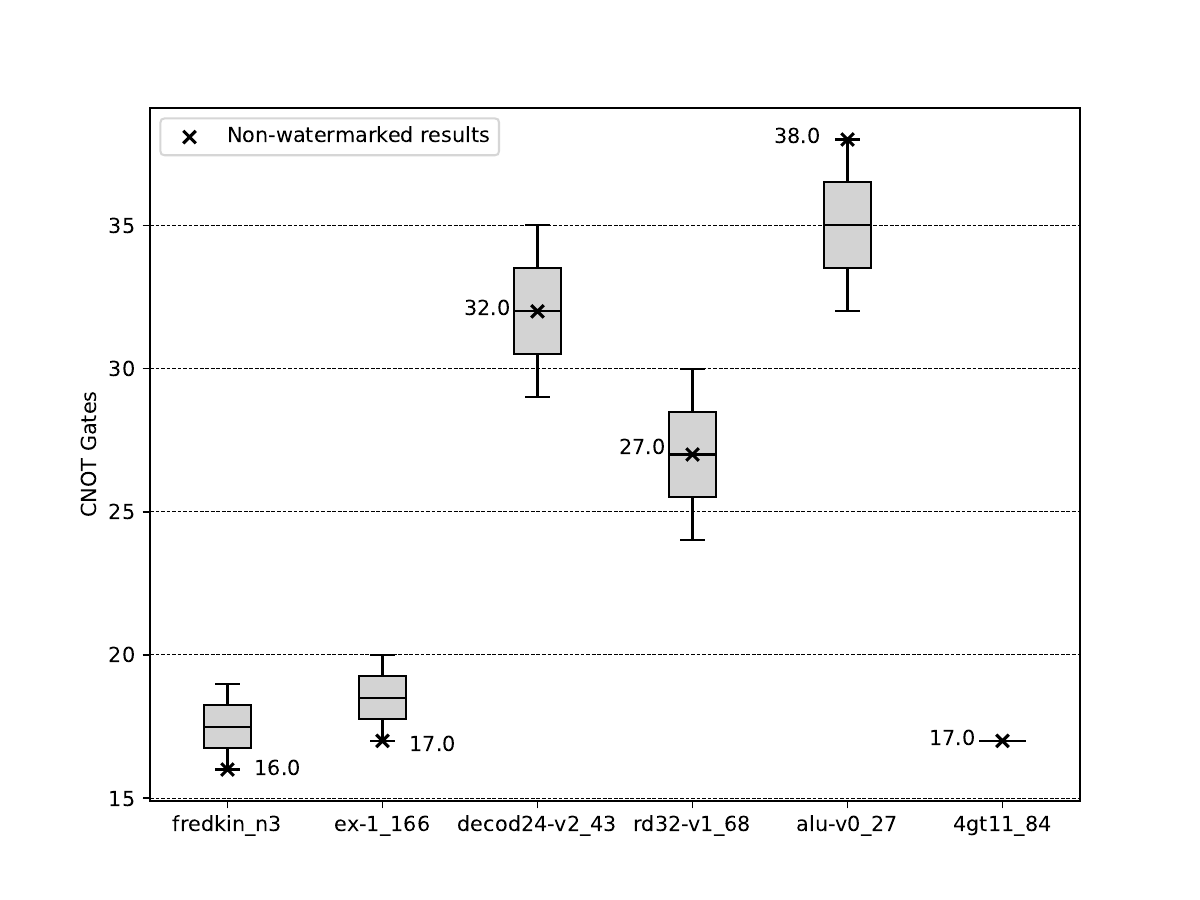}
   \label{f:Comparative2}
}
\\
\subfigure[PST.]{
   \includegraphics[width=3.3in]{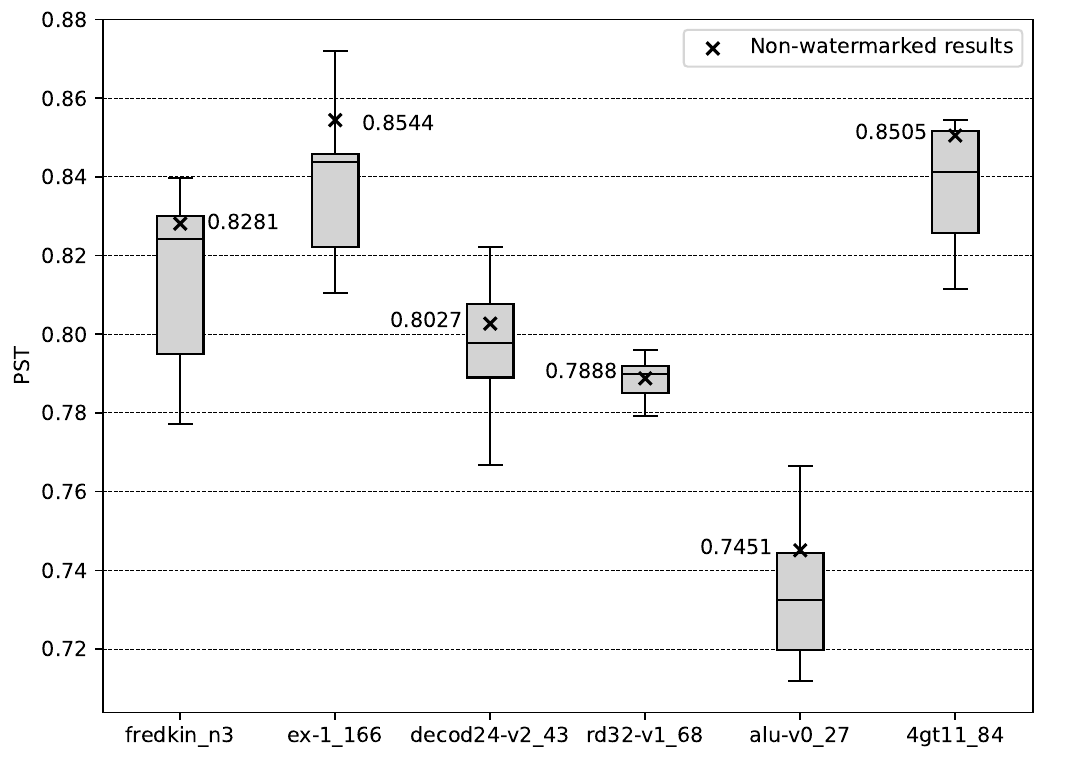}
   \label{f:Comparative3}
}
\hspace{-0.2in}
\subfigure[Synthesis time.]{
   \includegraphics[width=3.3in]{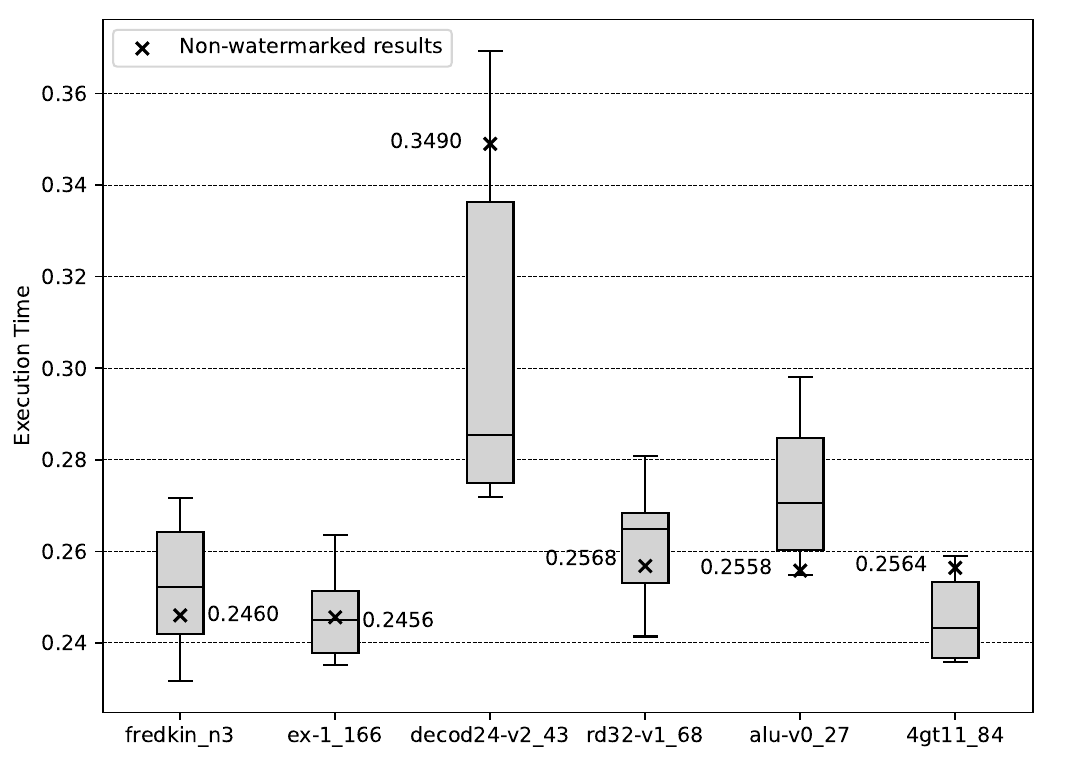}
   \label{f:Comparative4}
}
\vspace{-0.1in}
\caption{The design overhead for our multi-stage watermarking technique.}
\label{f:water_overhead_est}
\vspace{-0.2in}
\end{figure*}

\textbf{Circuit Benchmarks}. Table~\ref{details_benchmarks} presents six representative quantum circuits sourced from the prominent quantum computing benchmark, MQTBench~\cite{quetschlich2023mqtbench}. Our focus in this paper is on small-scale quantum circuits, given their sensitivity to watermarking insertion. These selected circuits range from 3 qubits to 5 qubits in logical qubit count. They comprise $8\sim22$ 2-qubit CNOT gates, resulting in a circuit depth spanning from 11 to 30. Each quantum circuit undergoes synthesis, mapping, and scheduling processes tailored for execution on an IBM NISQ machine simulator, FakeLagos.

\textbf{NISQ Machine}. We utilized a simulated fake backend of the IBM NISQ superconducting machine, FakeLagos, available in Qiskit~\cite{anis2021qiskit}. FakeLagos comprises seven physical qubits arranged in an ``H''-shape interconnection pattern. Figure~\ref{stage2_example} depicts the architecture of FakeLagos and the 2-qubit gate infidelity associated with each physical qubit on the machine.

\textbf{Circuit Synthesis}. We adopt the state-of-the-art quantum circuit compiler, Qiskit~\cite{anis2021qiskit}, in conjunction with the renowned optimizer, BQSKIT~\cite{bqskit}, to synthesize the simulated quantum circuit benchmarks. Customizing a BQSKit workflow for decomposition involves using the `ScanPartitioner'~\cite{younis2021qfast} option to divide the entire circuit into sub-circuits, each with a maximum size of three. Subsequently, each sub-circuit undergoes decomposition using the `QSearchSynthesisPass'~\cite{Davis:QCE2020} option, followed by reassembly using the `UnfoldPass' option. For the mapping stage, we employ the variation-aware qubit allocation policy~\cite{tannu2019not}, which optimizes qubit allocation to prioritize operations on qubits and links with higher fidelity. During the routing and optimization phase, we use the `SabreSwap'~\cite{li2019tackling} option to introduce SWAP gates, ensuring compatibility with the connectivity of the NISQ machine. Additionally, the `FullAncillaAllocation' option is employed to allocate all idle physical qubits as ancillary resources. In scheduling, `ALAPScheduleAnalysis' is used to schedule the end time of instructions as late as possible.

\section{Evaluation and Results}
\label{s:result}

\begin{table*}
\centering
\caption{The comparison between our multi-stage watermarking technique and our two baselines.}
\setlength{\tabcolsep}{3pt}
\begin{tabular}{|c||c|c|c|c||c|c|c|c|c||c|c|c|c|c|} 
\hline
\multirow{3}{*}{Benchmark} &\multicolumn{4}{c|}{non-watermarked} &\multicolumn{5}{c|}{\cite{Saravanan:ISQED2021}} &\multicolumn{5}{c|}{\textbf{Ours}} \\ \cline{2-15}
 & Circuit & 2-qubit    & \multirow{2}{*}{$\Delta$} & \multirow{2}{*}{PST}  & Circuit & 2-qubit    & \multirow{2}{*}{$\Delta$} & \multirow{2}{*}{PST} & \multirow{2}{*}{PPA}  & Circuit & 2-qubit    & \multirow{2}{*}{$\Delta$} & \multirow{2}{*}{PST} & \multirow{2}{*}{PPA}\\ 
 & Depth   & Gate \#    &   &                       & Depth   & Gate \# &  &  &    & Depth   & Gate \# &  &   &\\\hline\hline

fredkin\_n3                         & 31            & 16        & 1.55E-15   & 0.8281      & 40            & 21        & 2.21E-06   & 0.8032 & 0.25 & 32            & 16        & 1.78E-15   & 0.8135  & 0.0535\\ 
\hline
ex-1\_166                           & 34            & 17        & 6.66E-16   & 0.8544      & 31            & 16        & 5.49E-06   & 0.8251 & 0.25 & 36            & 18        & 1.78E-15   & 0.8402  & 0.0357\\ 
\hline
decod24-v2\_43                      & 63            & 32        & 2.11E-15   & 0.8027      & 69            & 35        & 4.29E-06   & 0.7231 &0.0625 & 63            & 32        & 2.78E-15   & 0.7963  & 0.0208\\ 
\hline
rd32-v1\_68                         & 50            & 27        & 1.44E-15   & 0.7888      & 58            & 30        & 1.55E-06   & 0.7523 & 0.0625& 53            & 26        & 3.33E-15   & 0.7884  & 0.0105\\ 
\hline
alu-v0\_27                          & 64            & 38        & 1.55E-15   & 0.7451      & 74            & 53        & 2.24E-06   & 0.5231  & 0.0156& 63            & 35        & 3.33E-15   & 0.7342  & 0.0052\\ 
\hline
4gt11\_84                           & 34            & 17        & 1.22E-15   & 0.8505      & 32            & 17        & 8.98E-06   & 0.8098 &0.25 & 38            & 17        & 1.33E-15   & 0.8373  & 0.0572\\
\hline
\end{tabular}
\label{t:water_all_results}
\vspace{-0.2in}
\end{table*}

\textbf{Overhead Estimation}. We assessed various types of design overhead introduced by our multi-stage watermarking technique by analyzing differences in the circuit depth, the number of 2-qubit CNOT gates, the PST values, and the synthesis time, as illustrated in Figure~\ref{f:water_overhead_est}. To provide a comprehensive evaluation, we employed random sampling to generate possible signatures and inserted them into the quantum circuits to evaluate design overhead. We considered the signature containing all `a's, all `c's, and no `e' as the non-watermarked baseline. Across the four plots, the x-axis categorizes different quantum circuit benchmarks, chosen to demonstrate a diverse range of watermarking effects. Each plot's y-axis corresponds to a specific circuit characteristic: circuit depth, CNOT gates, PST, and synthesis time. The `$\times$' symbol denotes values for non-watermarked results. Generally, compared to non-watermarked circuits, those watermarked by our technique exhibit increased circuit depth, a higher number of CNOT gates, reduced PST, and longer synthesis time.
\begin{itemize}[leftmargin=*, nosep, topsep=0pt, partopsep=0pt]
\item \textit{Circuit Depth}. As depicted in Figure~\ref{f:Comparative1}, the average depth of watermarked circuits experienced a maximum increase of 8\% and a decrease of up to 2\% compared to their non-watermarked counterparts. The anticipated increase in circuit depth can be attributed to the introduction of additional gates in stage-3 of our watermarking process. Conversely, despite the added gates in stage-3, a decrease in depth can occur due to the selection of different map positions in stage-2, which may reduce the overall circuit depth. This reduction may also result from fewer additional SWAP gates required during routing, effectively offsetting the depth increase caused by stage-3, leading to instances where the average depth decreased.

\item \textit{CNOT Gate $\#$}. As depicted in Figure~\ref{f:Comparative2}, our multi-stage watermarking technique typically led to an average increase or an equal count of 2-qubit CNOT gates across all quantum circuit benchmarks. However, a notable exception was observed in the benchmark `4gt11\_84', where the count of CNOT gates remained unchanged at 17 in all sampling results, indicating a particularly successful application of our multi-stage watermarking to this benchmark.

\item \textit{PST}. As depicted in Figure~\ref{f:Comparative3}, PST serves as a crucial metric for measuring the fidelity of watermarked quantum circuits. It was observed that our watermarking method resulted in a maximum average PST decrease of 1.7\% compared to the non-watermarked method, with the maximum PST in all benchmarks exceeding that of the non-watermarked method.

\item \textit{Synthesis Time}. As illustrated in Figure~\ref{f:Comparative4}, the average synthesis time for most watermarked benchmarks was moderately higher than that for non-watermarked methods, indicating that our watermarking technique can be efficiently implemented. In the case of `4gt11\_84', only a slight increase in execution time was observed, attributed to our watermarking technique's minimal impact on the CNOT gate count for this benchmark.
\end{itemize}

\textbf{Comparing Against Previous Work}. Our multi-stage watermarking technique (\textbf{Ours}) is compared against two baselines (non-watermarked baseline and~\cite{Saravanan:ISQED2021}) in Table~\ref{t:water_all_results}. In comparison to our non-watermarked baseline, across all quantum circuit benchmarks, our method yields a 3\% increase in circuit depth, a 2\% reduction in the number of CNOT gates, and a 1\% decrease in PST values, while achieving a probabilistic proof of authorship of 0.0304. In contrast, the previous watermarking technique~\cite{Saravanan:ISQED2021} results in an average increase of 10\% in circuit depth and a 16\% increase in CNOT gates, alongside a notable reduction of 19\% in PST values, with a probabilistic proof of authorship of 0.1484. We have the following observations:
\begin{itemize}[leftmargin=*, nosep, topsep=0pt, partopsep=0pt]
\item \textit{Circuit Depth and CNOT gate \#}. Regarding circuit depth and CNOT gate count, our technique introduces a smaller design overhead. Notably, for the `fredkin\_n3' and `4gt11\_84' benchmarks, our method exhibits a slight increase in circuit depth compared to the non-watermarked method but matches the 2-qubit CNOT gate count. This minimal depth increase may be attributed to stage 3 without significantly impacting the overall circuit complexity. For `ex-1\_166', our method results in a marginal increase in both depth and CNOT count compared to the non-watermarked method, yet it remains more efficient than the previous method, suggesting improvements over the prior watermarking technique~\cite{Saravanan:ISQED2021}.

\item \textit{$\Delta$ and PST}. The distance ($\Delta$) between the original unitary matrix and the decomposed unitary matrix of a quantum circuit is marginally higher in our method compared to the non-watermarked method, yet consistently lower than the previous method~\cite{Saravanan:ISQED2021}. Our method maintains $\Delta$ at the $1e$-$15$ level, while the previous technique~\cite{Saravanan:ISQED2021} increases $\Delta$ to the $1e$-$6$ level. Our method demonstrates a nuanced impact on PST, and achieves higher PST values than the prior watermarking scheme~\cite{Saravanan:ISQED2021} across various benchmarks. For instance, the `alu-v0\_27' benchmark shows a decrease of approximately 1\% in PST compared to the non-watermarked method, attributed to the additional complexity introduced by watermarking.

\item \textit{Tradeoff between design overhead and PPA}. Overall, our method consistently performs on par or better in terms of circuit depth, CNOT gates, $\Delta$s, and PST values, compared to the previous method~\cite{Saravanan:ISQED2021}. However, compared to the prior technique~\cite{Saravanan:ISQED2021}, our multi-stage watermarking technique demonstrates a much lower PPA.
\end{itemize}

\section{Conclusion}
\label{s:con}

In this paper, we have presented a novel multi-stage watermarking scheme tailored specifically for quantum circuits. By integrating additional constraints across various synthesis stages, our approach enhances the robustness and effectiveness of IP protection in quantum computing. Compared to the state-of-the-art watermarking technique, our multi-stage watermarking approach demonstrates, on average, a reduction in the number of 2-qubit gates by 16\% and circuit depth by 6\%, alongside an increase in the fidelity of watermarked circuits by 8\%, while achieving a 79.4\% lower probabilistic proof of authorship.

\bibliographystyle{IEEEtran}
\bibliography{quantum}
\end{document}